\documentclass{aadebug}
\corr{0309027}{87}
\begin{document}

\runningheads{Kazutaka Maruyama et al.}{Timestamp Based Execution Control for C and Java Programs}

\title{Timestamp Based Execution Control for C and Java Programs}

\author{
Kazutaka~Maruyama\addressnum{1}\comma\extranum{1},
Minoru~Terada\addressnum{2}\comma\extranum{2}
}

\address{1}{
Dept. of Mechano-Informatics,
Grad. School of Information Science and Technology,
The University of Tokyo,
Japan
}

\address{2}{
Dept. of Information and Communication Engineering,
The University of Electro-Communications,
Japan
}

\extra{1}{E-mail:kazutaka@acm.org}
\extra{2}{E-mail:terada@ice.uec.ac.jp}

\pdfinfo{
/Title (Timestamp Based Execution Control for C and Java Programs)
/Author (Kazutaka Maruyama, Minoru Terada)
}


\begin{abstract}
Many programmers have had to deal with an overwritten variable
resulting for example from an aliasing problem.
The culprit is obviously the last write-access to that memory 
location before the manifestation of the bug.
The usual technique for removing such bugs starts with the debugger
by (1) finding the last write and (2) moving the control point
of execution back to that time by re-executing the program from 
the beginning.  
We wish to automate this.
Step (2) is easy if we can somehow mark the last write found in 
step (1) and control the execution-point to move it back to this time.

In this paper we propose a new concept, \textit{position}, that is,
a point in the program execution trace, as needed for step (2) above.
The position enables debuggers to automate the control of program
execution to support common debugging activities.
We have implemented position in C by modifying GCC and in Java with
a bytecode transformer.  Measurements show that position can
be provided with an acceptable amount of overhead.
\end{abstract}




\section{Introduction}\label{sec:intr}
Studies about
formal system specification with mathematical notation\cite{ince-z},
automatic testing by semantics\cite{meyer}, and so on,
have continued for long years.
However,
these projects still evolve and cannot be used for real problems
and are not enough to exterminate all bugs.
Programmers must debug programs by hand.

In debugging,
we usually use debuggers to observe the behavior of programs.
Debuggers have various functions supported by hardware
and operating system\cite{howdebuggerswork},
but the functions offer too low-level commands.
Operations that programmers want to do in debugging
are more abstract than raw commands of debuggers,
and they must break down their operations into them.
Programmers have to waste energy thinking about how to use debugger commands
while they examine the behavior of programs
because of a gap between what debuggers can do
and what programmers need.
On the other hand,
there are some patterns of the operations
which programmers want debuggers to do.
To automate them is useful for efficient debugging.

In this paper,
we propose a new idea \textit{position}
as a base technique of the execution control
useful for automating some typical debugging operations
that programmers want to do.
In order to implement it,
we introduce a counter \textit{timestamp}
which increases whenever the control point jumps backward
and we insert the codes for updating timestamp
into programs to be debugged.
We describe the implementation details for C and Java programs.
Overhead measurements of programs with the updating codes
are also included.

The rest of this paper is organized as follows.
Section \ref{sec:pos} proposes the notion of ``position''
and describes the advantages of its applications.
Section \ref{sec:implc} and \ref{sec:implj}
describe the implementation details for C and Java respectively,
along with the result of overhead measurements.
Section \ref{sec:vs} describes another representation of position
without timestamp and the difference between the two.
Section \ref{sec:relw} discusses the relevance to other works.
Section \ref{sec:conc} and \ref{sec:futu}
describe conclusion and future work.


\section{Position: New Idea for Execution Control}\label{sec:pos}
Figure \ref{fig:layer} shows the structure of our proposal.
In this section,
we first propose the idea of the position
and introduce ``timestamp'' as its base.
Next, we implement a simple application of the position,
``dynamic breakpoint''.
Finally, we describe three applications of the position,
``bookmarking positions'', ``reverse watchpoint''
and ``binary search method''.

\begin{figure}
\centering
\includegraphics{layer}
\caption{Structure of our proposal}
\label{fig:layer}
\end{figure}

\subsection{Timestamp and Position}\label{subsec:pos-ts}
We introduce a new idea, \textit{position},
in order to specify one point in the program trace,
the series of statements executed in order of time.\footnote{
The \textit{point} is really \textit{one statement} of the source code.}
The position introduces an absolute coordinate in program traces
and indicates a target point of the execution control.

In debugging, the control of program execution used so far
is based on static information such as line numbers in source codes
and cannot express the position
because the backward jumps of the control point
may cause multiple executions of one statement.
To distinguish them from each other,
we introduce a new counter into debuggees.
We call the counter \textit{timestamp},
which increases whenever the control point jumps backward.
The position is expressed as the pair of
the line number and the value of the timestamp.

\begin{figure}
\centering
\includegraphics{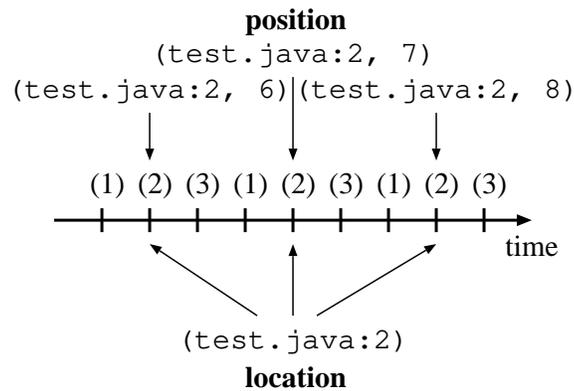}
\caption{Location and position in program trace}
\label{fig:position}
\end{figure}

An example code with a loop structure
is shown in figure \ref{fig:ts-sample}.
There are multiple appearances of three lines (1) to (3)
in the program trace (figure \ref{fig:position}).
We call the static point of execution expressed as the line number
\textit{location}.

\begin{figure}
{\small
\ \hrule \ 
\begin{verbatim}
        :
(1) while(i < a){
(2)   i += b;
(3) }
        :
\end{verbatim}
\ \hrule \ }
\caption{Code with a loop structure}
\label{fig:ts-sample}
\end{figure}

Timestamp increases whenever the loop body is repeated,
so the pair of the line number and the value of timestamp
expresses the position, the dynamic point of execution.
Using the value of timestamp,
we can distinguish each of multiple appearances of a location
in the trace.
For Java programs,
timestamp should increase at the following cases:
\begin{itemize}
\item entrance and exit of method invocations,
\item loop body,
\item exception.
\end{itemize}

Position could be expressed as the whole history of debugger commands,
rather than the pair.
We discuss this topic in section \ref{sec:vs}.

\subsection{Dynamic Breakpoint}\label{subsec:pos-dbp}
We here describe how to move the control point to a position.
We propose a new breakpoint facility,
\textit{dynamic breakpoint},
to be set at a position, rather than at a location.\footnote{
We call the normal breakpoint \textit{static breakpoint}.}
Because the implementation of it needs the existing breakpoints
we get support from debuggers.

A simple implementation of dynamic breakpoint
could use ``conditional breakpoint'' of debuggers.
We show an instruction example
with the notation of GDB\cite{gdb}.
For example,
if we want to stop the debuggee at the position,
\texttt{(test.java:2, 8)},\footnote{
Debugger already stopped at the position and recorded the timestamp
in a previous run.
If second execution path is unexpectedly different from first,
the recorded position becomes meaningless and
there is no way the debuggee knows such change by itself.
Discussion about non-deterministic debuggees are in section \ref{sec:futu}.}
we instruct Java debugger, JDB, as follows:\footnote{
JDB does not support conditional breakpoints yet.}
\begin{verbatim}
break test.java:2 if Timestamp.ts = 8
\end{verbatim}

This implementation has performance problem,
since many context switchings may occur
in order to evaluate the given conditions.
We will show more effective implementation later.

\subsection{Applications of Position}\label{subsec:pos-app}
We describe three applications of position.
We assume that debuggees are deterministic including
the execution environments;
we discuss non-deterministic cases in section \ref{sec:futu}.

\subsubsection{Bookmarking Positions}\label{subsubsec:pos-app-mark}
When we know where the cause of the bug is only roughly,
we use breakpoints to move the control point before the target position,
and then use a step execution repeatedly.
If a programmer who walks through a large program
mistakenly pass over the desired position,
he must recall and replay all the commands
he had given from the beginning.

Bookmarking positions is like a mountaineer
placing anchors for his rope as he goes along.
We think that the behavior of the debuggee is correct
as far as here,
then we bookmark the position by using a dynamic breakpoint.
If we mistake something later,
we can go back to the lost anchor point
before things went wrong.

Furthermore,
if a programmer annotates the position
as its identifier instead of the ID number,
he would remember the position easily.
For example,
the comment might be:
``Just read a right brace; the parser is about to process
a compound statement.''

\subsubsection{Reverse Watchpoint}\label{subsubsec:pos-app-rw}
Suppose that a program allocates memory dynamically.
If the program writes beyond the range of allocated area,
it may destroy the header information of it used by
allocator functions \texttt{malloc} and \texttt{free}.
But the destruction operation itself does not cause
the bug manifestation immediately,
and the result is usually manifested much later.
Similarly,
if an object is unexpectedly pointed
from two different references (called \textit{aliasing}),
the content of the object could be destroyed.

It is difficult to fix these bugs
which are caused by writing an invalid value to a variable unexpectedly
because their manifestation occurs later than the write.
To catch these invalid writes,
debuggers provide data access breakpoint facility,
\textit{watchpoint} in GDB,
which traps all write accesses to a certain variable.
First, in debugging,
we examine which variable are invalid.
Second, we look for the operation which destroyed it
by using watchpoint.

When we use watchpoint,
the debuggee stops many times
and we examine all the output
to know whether the write access is relevant to the bug.
Such work requires too much time for us
to concentrate our attention on the whole debugging session
and to find a sign of the bug to be found.

On the other hand,
the last write to the variable obviously causes the bug manifestation.
We do not know which write is the last one until the bug manifests.
To know the write,
we must set a watchpoint at the variable,
re-start the debuggee with counting stops by the watchpoint
until the control point reaches where bug manifests,
and re-start it again to go back to the last write.
We are going to automate this procedure.

We propose the ``reverse watchpoint'' facility
which automatically moves the control point of the debuggee
to the last write to a certain variable.
This new debugger command takes a variable name
to be observed as its argument and does such control movement.
Using the reverse watchpoint,
programmers do not have to care about each stop by watchpoint
and can concentrate their attention on more intelligent work in debugging.

Reverse watchpoint is easily implemented
by using the dynamic breakpoint described above
and existing debugger.
The procedure is as follows.

\begin{figure}
\centering
\includegraphics{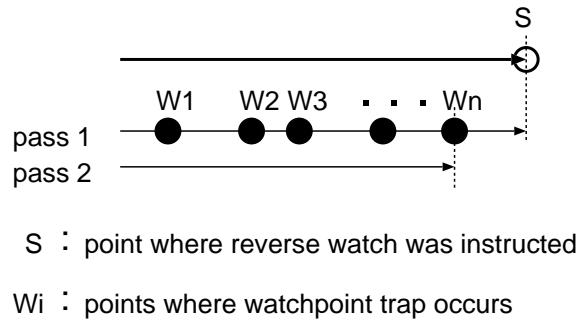}
\caption{Procedure of reverse watchpoint}
\label{fig:revwatch}
\end{figure}

\begin{enumerate}
\item
Set a dynamic breakpoint at the position
where reverse watchpoint is instructed
(\textbf{S} in figure \ref{fig:revwatch}).
\item
Pass 1:
\begin{enumerate}
\item
set a normal watchpoint at the target variable
and re-start the debuggee.
Whenever it stops by the trap of the watchpoint,
collect the information for its position
(actually the value of the timestamp)
in order to mark the position
\textbf{W1} to \textbf{Wn} in figure \ref{fig:revwatch}.
\item
Repeat the collection
until the control point reaches the position \textbf{S}.\footnote{
We can know it by stopping at the dynamic breakpoint
set at step 1.}
\end{enumerate}
\item
Pass 2:
\begin{enumerate}
\item
go back to the beginning again,
set another dynamic breakpoint at the most recent position \textbf{Wn},
and re-start it.
\item
The program stops at the target position \textbf{Wn}.
\end{enumerate}
\end{enumerate}

This procedure corresponds to
creating dynamic slicing\cite{dynamic-slicing} manually.
The advantage of our proposal is
that control point of execution can be moved at
interesting positions where a variable is assigned
without problems about ``unconstrained pointers''\cite{slicing-survey},
such as those found in C.
Slicing techniques are based on analysis of source codes,
but watchpoint of debuggers receive support from hardware
for fully precise information of assignments.

\subsubsection{Binary Search Method for Locating Bugs}\label{subsubsec:pos-app-bs}
We propose a binary search method for locating bugs
using timestamp as a facility for automatically locating
the position where a condition becomes false at the first time.

Suppose that a program constructs a doubly linked list and
we want to find the position
where the consistency of the list is lost.
We can do this more easily by using timestamp
as an index of the binary search method.
A driver program which implements the method
should control a debuggee as follows:
\begin{enumerate}
\item
move the control point to the position where the timestamp value
is the middle of the left end
(initially, the beginning of the execution)
and the right end
(first, the end),
\item
evaluate the condition,
and
\begin{itemize}
\item
if it is true then go to the position
between the current position and the right end,
and narrow the target area by dealing with
the new position as the new left end,
\item
if it is false then go back to the position between
the left end and the current,
and narrow the target area by dealing with
the new position as the new right end,
\end{itemize}
\item
repeat above.
\end{enumerate}

Inserting assertions at various source lines
and evaluating them repeatedly
might seem to achieve the same effect as ours.
But our method has two advantages.
First,
\texttt{assert} must be inserted manually
at all the locations where it may be necessary.
The binary method does not have to do.
Second,
the inserted \texttt{assert}s cause
condition evaluations at each call.
Using our method,
the maximum number of the examinations is $\log n$
such that $n$ is the timestamp value at the end of the execution.
So, if the condition is complex,
the performance of the method may be better than the one
of \texttt{assert}.

This method is already proposed
by Tolmach and Appel\cite{sml-debugger}.
We take it
from the world of functional programming
to one of procedural programming.

Because binary search method requires the condition
to be monotonous (or at least become false at the end),
we might have to repeat the process
in order to arrive the moment of true bug.
Suppose the following scenario:
the true bug is a temporal invalid value of a variable in a condition C1;
the invalid value is propagated to another variable in a condition C2;
C1 recovers the correct state;
C2 causes the crash.
If we use C1 to test the debuggee,
we cannot find the moment C1 becomes invalid
because the condition is not monotonous.
But C2, the direct cause of the crash,
leads us to the moment when it becomes invalid.
After this step,
we use C1 to arrive at the true bug.


\section{Implementation for C}\label{sec:implc}
We need to modify target program
to include codes for updating timestamp.

\subsection{Discussion about Target of Transformation}\label{subsec:implj-gcc}
We have implemented the transformer in intermediate code level.
There might be three levels of C program transformation:
\begin{enumerate}
\item source code level,
\item intermediate code level,
\item assembly code level.
\end{enumerate}

The first one has the advantages of 
independence from hardware platforms,
operating systems, and compilers.
However,
codes which programmers see in debugging
are different from ones they wrote
and the implementation is slightly difficult
because the transformer have to analysis the output of C preprocessor.

The advantage of the second one
is to be independent of target architectures
if the compiler supports that platform,
while the disadvantage is to be dependent on a certain compiler.

The target of the third level is assembly code emitted by compilers.
We first implemented a simple transformer
in this level\cite{kazutaka-revfunc},
because it is the easiest method,
but is dependent on target architectures
and lacks the portability.

GNU C Compiler (GCC)\cite{gcc} is chosen as the target compiler
because it is used in various platforms.
GCC generates the intermediate code,
called Register Transfer Language (RTL),
from the source code
and we modified a part of GCC to insert the codes of increment of timestamp
at the RTL generation stage.
The target of our implementation is GCC-2.95.2
which was the latest release of GCC at the time.
The details of the modification is described
in section \ref{subsec:implc-gcc}.

When compiled, debuggees are inserted
the macro \texttt{INC\_TS} shown in figure \ref{fig:incts}
at proper locations.

\begin{figure}
\centering
\ \hrule \ 
\begin{verbatim}
int timestamp = -1;
int ref = -1;

void brake(void){}

#define INC_TS if(++timestamp == ref) brake();
\end{verbatim}
\ \hrule \ 
\caption{Codes for updating timestamp}
\label{fig:incts}
\end{figure}

\subsection{Implementation of Dynamic Breakpoint for C}\label{subsec:implc-dbp}
\begin{figure}
\centering
\includegraphics{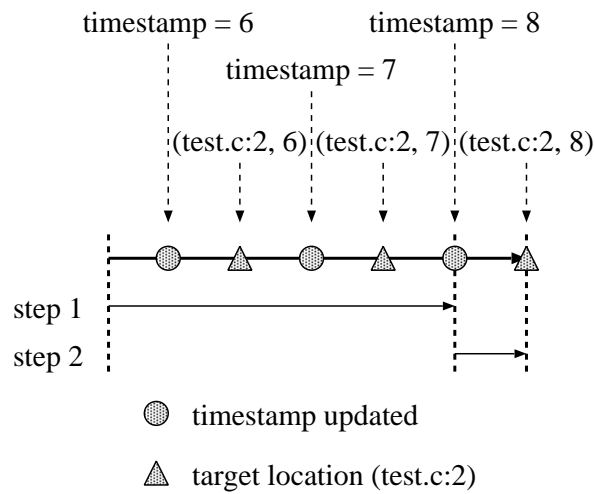}
\caption{Procedure of dynamic breakpoint}
\label{fig:dbp}
\end{figure}

The implementation of dynamic breakpoint
described in section \ref{subsec:pos-dbp}
has performance problem.
Conditional breakpoint evaluates the given condition
whenever the control point reaches the location
shown as $\triangle$ in figure \ref{fig:dbp},
so the execution usually slows down seriously.

The implementation with the least overhead (only 2 breaks)
is as follows.

\begin{description}
\item[Step 1]
Prior to the execution,
set a static breakpoint in the function \texttt{brake()},
assign the target value of timestamp
to the debuggee's variable \texttt{ref}.
Start the execution
until it stops by the breakpoint
when the value of timestamp reaches \texttt{ref}.
\item[Step 2]
Set another static breakpoint at the target location
and continue the execution.
\end{description}

\subsection{Modifications of GCC}\label{subsec:implc-gcc}
In GCC,
the function of the parser, \texttt{yyparse},
invokes its actions which generate RTL.
We modify GCC
so that it has a new command line option, \texttt{-pg2},
and emits the RTL for updating timestamp
if the option is given.
In C programs,
timestamp should increase at the following cases.\footnote{
Our implementation does not increase timestamp
when \texttt{longjmp()} is called.
It does not matter
unless \texttt{setjmp()} and \texttt{longjmp()} are called
in the same function,
which is usually expressed by \texttt{goto} statement.}

\begin{figure}
\centering
\includegraphics{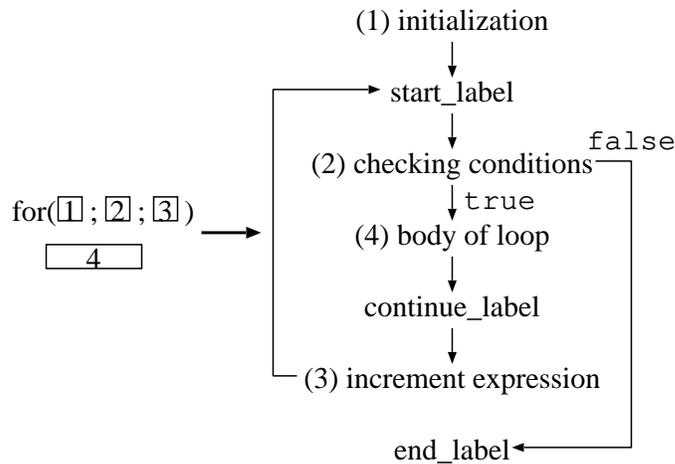}
\caption{Expansion of \texttt{for} statements}
\label{fig:loop-expand}
\end{figure}

\begin{description}
\item[loops]
The increment code is emitted
just after the label, \texttt{start\_label},
which is placed at the start of loops in \texttt{expand\_start\_loop}.
The label is the target of jumps from tails of loop bodies
and is emitted after the initialization of \texttt{for} statements
(figure \ref{fig:loop-expand}).
It works well in the cases of \texttt{while} statements
and \texttt{do/while} statements.\footnote{
Timestamp should increase
just after \texttt{continue\_label}
in terms of backward jumps.
But we do not choose this approach
for the unification of the implementations
for three kinds of loop statements.
Therefore one extra increment of timestamp occurs
at each loop structure,
but it is negligible.}
\item[\texttt{goto} statements]
The increment code is emitted
just before the invocation of\\
\texttt{expand\_goto\_internal}
in \texttt{expand\_goto}.
Jumps of which the destination label is forward
do not need to cause the increment of timestamp.
But we do not decide whether a jump is backward
for the sake of keeping the implementation simple.\footnote{
Too frequent updating of timestamp does not destroy
the consistency of position,
but causes a drop in the performance.}
GCC has its original features,
``nonlocal goto'' and ``computed goto''.
Our modified GCC does not regard these \texttt{goto}s
as the target of the increment of timestamp
because of the simple implementation.
\item[\texttt{return} statements]
\texttt{return}s with a return value
put it in the registers of CPU.
The increment code is emitted
just before the computation of arguments of \texttt{return}s
in \texttt{c\_expand\_return}
in order to prevent the value in the registers from being destroyed.
\item[entrance of function calls]
The code is emitted
just after the invocation of \texttt{store\_parm\_decls},
which registers the name and type of the arguments of the function call.
\item[exit of function calls]
This is relevant to the tail of \texttt{void} functions
without explicit \texttt{return}.
In order to prevent extra increments
after one at \texttt{return} statements,
the code is emitted
just before \texttt{return\_label}
which is emitted in epilogues of functions.\footnote{
\texttt{return\_label} is the destination of jumps
from \texttt{return} statements.}
\end{description}

Our modifications to GCC have around 70 lines.
We have confirmed the behavior of our GCC
on four platforms.
In order to port it to other platforms,
all we have to do is the modification
of a macro which is relevant to command line options
in a header file for each platform.

\begin{itemize}
\item i386-linux
\item alpha-linux
\item sparc-sunos4
\item sparc-sun-solaris2.8
\end{itemize}

Figure \ref{fig:addedcode} shows
the inserted codes on i386-linux and sparc-sunos4
in assembly code notation.
The codes are generated with options
for debugging, optimizations and timestamp.

\begin{figure*}[t]
\centering
\ \hrule \ 
\begin{verbatim}
incl timestamp                 sethi   %hi(_timestamp), %o1
movl timestamp,%eax            ld      [%o1+%lo(_timestamp)], %o0
cmpl ref,%eax                  add     %o0, 1, %o0
jne .L6                        st      %o0, [%o1+%lo(_timestamp)]
call brake                     sethi   %hi(_ref), %o1
                               ld      [%o1+%lo(_ref)], %o1
                               cmp     %o0, %o1
                               bne     L21
                               mov     0, %l2
                               call    _brake, 0
\end{verbatim}
\ \hrule \ 
\caption{Inserted assembly codes for i386-linux (left) and sparc-sunos4 (right)}
\label{fig:addedcode}
\end{figure*}

\subsection{Overhead Measurement of C Implementation}\label{subsec:implc-over}
Table \ref{tab:overheadc} shows the result of measurement
of runtime overhead.
The debuggee of this benchmark is gawk-2.15.6.
Overhead of other C applications is also measured (table \ref{tab:overheadc2}).
The benchmark ``empty loop'' has only one loop structure
whose body is empty and shows the worst case of overhead of our system,
because both increment of timestamp and comparing it to \texttt{ref} occur
whenever the loop is repeated.
The ``sed'' benchmark is sed-1.18.
Platforms in these tables are shown in table \ref{tab:platforms}.\footnote{
We regret to lose SPARC-1 platform and cannot know
its clock of CPU.}

The number of increments of timestamp is
500000002 in ``loop'', 85888810 in ``sed'' and
96817515 in ``gawk''.
The average time required for increment is about
$13.7$, $9.76$ and $16.5$ nanoseconds respectively.
We think it is acceptable for practical use.

\begin{table}
\centering
\caption{Overhead of gawk with timestamp system}
\label{tab:overheadc}
\hbox to\hsize{\hfil
\begin{tabular}{|l|r|r|r|}
\hline
GCC options & Intel-1 & Alpha & SPARC-1 \\
\hline\hline
-g -O & 2.57 & 4.64 & 17.44 \\
& (1.00) & (1.00) & (1.00) \\
\hline
-g -O -pg2 & 3.78 & 6.34 & 24.40 \\
& (1.47) & (1.37) & (1.40) \\
\hline
\multicolumn{4}{r}{seconds (ratio)}\\
\end{tabular}\hfil}
\end{table}

\begin{table}
\centering
\caption{Overhead of other C applications and platforms}
\label{tab:overheadc2}
\hbox to\hsize{\hfil
\begin{tabular}{|l|l|r|r|}
\hline
App. & GCC options & Intel-2 & SPARC-2 \\
\hline\hline
empty loop & -g -O & 2.762 & 1.012 \\
& & (1.00) & (1.00) \\
\cline{2-4}
& -g -O -pg2 & 9.632 & 10.112 \\
& & (3.49) & (9.99) \\
\hline\hline
sed & -g -O & 9.182 & 3.182 \\
& & (1.00) & (1.00) \\
\cline{2-4}
& -g -O -pg2 & 10.02 & 4.916 \\
& & (1.09) & (1.54) \\
\hline\hline
gawk & -g -O & 2.87 & 2.842 \\
& & (1.00) & (1.00) \\
\cline{2-4}
& -g -O -pg2 & 4.466 & 4.598 \\
& & (1.56) & (1.62) \\
\hline
\multicolumn{4}{r}{seconds (ratio)}\\
\end{tabular}\hfil}
\end{table}

\begin{table}
\centering
\caption{Platforms of C applications}
\label{tab:platforms}
\hbox to\hsize{\hfil
\begin{tabular}{|l|lr|l|}
\hline
\multicolumn{1}{|c|}{Platform} & \multicolumn{2}{c|}{Architecture} &
  \multicolumn{1}{c|}{OS} \\
\hline
Alpha   & Alpha   & 500MHz & Linux (glibc2) \\
Intel-1 & Celeron & 400MHz & Linux (glibc1) \\
Intel-2 & Celeron & 366MHz & Linux (glibc2) \\
SPARC-1 & SPARC   & N/A    & SunOS-4 \\
SPARC-2 & UltraSPARC & 500MHz & Solaris-8 \\
\hline
\end{tabular}\hfil}
\end{table}

Table \ref{tab:overheadcsize} shows the total size of files of
each benchmark.
The increases are acceptable.

\begin{table}
\centering
\caption{Increase of file size of C applications}
\label{tab:overheadcsize}
\hbox to\hsize{\hfil
\begin{tabular}{|l|l|r|r|}
\hline
App. & GCC options & Intel-2 & SPARC-2 \\
\hline\hline
empty loop & -g -O & 1144 & 2160 \\
& & (1.00) & (1.00) \\
\cline{2-4}
& -g -O -pg2 & 1248 & 2416 \\
& & (1.09) & (1.12) \\
\hline\hline
sed & -g -O & 41030 & 55263 \\
& & (1.00) & (1.00) \\
\cline{2-4}
& -g -O -pg2 & 52430 & 77215 \\
& & (1.28) & (1.40) \\
\hline\hline
gawk & -g -O & 140046 & 169310 \\
& & (1.00) & (1.00) \\
\cline{2-4}
& -g -O -pg2 & 169414 & 223078 \\
& & (1.21) & (1.32) \\
\hline
\multicolumn{4}{r}{bytes (ratio)}\\
\end{tabular}\hfil}
\end{table}


\section{Implementation for Java}\label{sec:implj}
From the experience of C,
we have chosen the implementation using bytecode transformation.
We considered four levels to implement the transformer for Java
before the decision:
\begin{enumerate}
\item compilers (source code level),
\item class files (bytecode level),
\item virtual machines (bytecode execution level),
\item Java Platform Debugger Architecture\cite{jpda}
(debugger interface level).\footnote{
Java Platform Debugger Architecture (JPDA) is included in JDK.
Java debugger JDB also uses it.}
\end{enumerate}

The first and the third are excluded
because of the same reason (portability) as the case of C.
The fourth is very portable
but we excluded this approach
because of the expected high overhead.

Although bytecode can be regarded as assembly code in the case of C,
the format is established
by Java virtual machine specification\cite{javavmspec},
and it is independent of target architectures.

\subsection{List of Bytecodes which Cause Increment of Timestamp}\label{subsec:implj-bc}
As explained in section \ref{subsec:pos-ts},
timestamp must be updated at several types of program structure.

\begin{description}
\item[entrance and exit of method invocations]
No bytecodes correspond with the entrance of method invocations.
The top of method body is used instead.
\texttt{return} bytecodes (opcode: 172--177)
correspond with exit of methods.\\
Bytecodes which invokes methods such as \texttt{invokevirtual}
could be used instead of the beginning of method body.
We selected the latter,
because the amount of added codes is less than the former.
Another advantage of this choice
is that the timestamp overhead is only added
to the methods of the modified class.
There will be no overhead for calling non-modified classes
(such as ones in system library).
\item[branches]
\texttt{ifeq} (153) to \texttt{if\_acmpne} (166),
\texttt{ifnull} (198) and \texttt{ifnonnull} (199).
\item[goto]
\texttt{goto} (167) and \texttt{goto\_w} (200).
\item[other jumps]
\texttt{jsr} (168), \texttt{ret} (169) and \texttt{jsr\_w} (201).
\item[exceptions]
\texttt{athrow} bytecode throws exceptions,
but there exists certain exceptions
which is not \texttt{athrow}ed explicitly
(such as NullPointerException).
Instead the codes of increment of timestamp are inserted
into the entry of the \texttt{catch} block.
\end{description}

Our transformer inserts a bytecode for updating timestamp,
just before bytecodes described above.
The inserted one is only \texttt{invokestatic}
followed by the index number of Methodref tag
which indicates \texttt{Timestamp.inc()} in constant pool.
Bytecodes of \textit{branches}, \textit{goto}s and \textit{other jumps}
described above have an operand which designates an offset
to its target address.
The update of timestamp is inserted when the operand has a negative value,
i.e. backward jump.
The implementation of \texttt{Timestamp} class
at present is shown in figure \ref{fig:ts-class}.

\begin{figure}
{\small
\ \hrule \ 
\begin{verbatim}
public final class Timestamp{
  private static long ts, ref;

  static{
    ts = 0;
    ref = 0;
  }

  public static void inc(){
    if(++ts == ref) brake();
  }

  private static void brake(){}
}
\end{verbatim}
\ \hrule \ }
\caption{\texttt{Timestamp} class}
\label{fig:ts-class}
\end{figure}

\subsection{Bytecode Transformer}\label{subsec:implj-conv}
Our bytecode transformer which transforms Java class files
is written in Java using Bytecode Engineering Library\cite{bcel} (BCEL)
and has around 160 lines.

It is not necessary to transform all class files of a program:
it is possible to do only those classes the user considers suspicious.
This reduces the overhead significantly.

\subsection{Overhead Measurement of Java Implementation}\label{subsec:implj-over}
We show the results of measurement of
runtime overhead and increase of size of transformed class files.
The target class files include an empty loop
and benchmark programs of SPEC JVM98\cite{specjvm98}.
We run them under JDK-1.4.0 on Linux PC
(Pentium III 733MHz, 640MB memory, and Linux-2.4.7).

We ran each benchmark seven times
and found the means of five results except the best and the worst.
The result is shown in table \ref{tab:overhead}.
In the case of empty loop,
whenever the loop is repeated,
the method invocation of \texttt{Timestamp.inc()} occurs.
This benchmark shows the worst case of overhead of our system.
Our implementation at present slows down around 4 times.
For other benchmarks
overheads are around 1.5 to 2 times of slowing down.
We think it is acceptable for practical use.
When the implementation of reverse watchpoint completed,
its overhead would be the sum of two:
timestamp system overhead which is
around 1.5 to 2 for each of two pass
and watchpoint overhead which JDB produces.
We may estimate the overhead of reverse watchpoint
to be less than around 4 times of slowing down in most cases.
Note that \texttt{\_227\_mtrt} benchmark is a multi-threaded program
and we added \texttt{synchronized} to \texttt{Timestamp.inc()} method
only for this benchmark.
So the overhead is heavier than others.

\begin{table*}[t]
\caption{Overhead of Java Programs with Timestamp System}
\label{tab:overhead}
\hbox to\hsize{\hfil
\begin{tabular}{|l|rr|rr|}
\hline
Benchmark & \multicolumn{2}{c|}{Original} &
 \multicolumn{2}{c|}{Backward Jumps}\\
\hline
\hline
empty loop & 12.774 & (1.00) & 56.416 & (4.42)\\
\hline
\_201\_compress & 1.4006 & (1.00) & 2.7754 & (1.98)\\
\_202\_jess & 0.2126 & (1.00) & 0.271 & (1.27)\\
\_209\_db & 0.412 & (1.00) & 0.433 & (1.05)\\
\_222\_mpegaudio & 0.1826 & (1.00) & 0.2792 & (1.52)\\
\_227\_mtrt & 0.363 & (1.00) & 0.9038 & (2.49)\\
\_228\_jack & 0.603 & (1.00) & 0.7212 & (1.19)\\
\hline
\multicolumn{5}{r}{seconds (ratio)}\\
\end{tabular}\hfil}
\end{table*}

Table \ref{tab:filesize} shows the total size of files of each benchmark.
The increase of empty loop benchmark whose file size is very small
and that of \texttt{\_201\_compress} benchmark are a little large,
but others increase little.

\begin{table*}[t]
\caption{Increase of Java class file size}
\label{tab:filesize}
\hbox to\hsize{\hfil
\begin{tabular}{|l|rr|rr|}
\hline
Benchmark & \multicolumn{2}{c|}{Original} &
 \multicolumn{2}{c|}{Backward Jumps}\\
\hline
\hline
empty loop & 276 & (1.00) & 328 & (1.19)\\
\hline
\_201\_compress & 14443 & (1.00) & 18640 & (1.29)\\
\_202\_jess & 396536 & (1.00) & 407240 & (1.03)\\
\_209\_db & 10156 & (1.00) & 10588 & (1.04)\\
\_222\_mpegaudio & 120182 & (1.00) & 124438 & (1.04)\\
\_227\_mtrt & 859 & (1.00) & 920 & (1.07)\\
\_228\_jack & 132516 & (1.00) & 138109 & (1.04)\\
\hline
\multicolumn{5}{r}{bytes (ratio)}\\
\end{tabular}\hfil}
\end{table*}


\section{Debugger Command History as a Position}\label{sec:vs}
We chose \textit{timestamp representation},
the pair of line number and timestamp, for position.
Another one is \textit{command-history representation}.
When we arrive at a certain position in a debugging session,
the whole history of debugger commands enables us
to come back to the position by re-execution from the beginning.
The advantages of our representation are as follows.

\begin{description}
\item[Total order]
Timestamp representation is totally ordered
and any position can be compared to each other in order of time,
so we can do binary search method described
in section \ref{subsubsec:pos-app-bs}.
\item[Uniqueness]
Our choice gives a unique representation for a position
while there may be many command histories
leading to a position.
This enables a 1-to-1 correspondence
between a position and a bookmark for it.
\item[Performance]
In most cases,
playing back debugger commands requires high overhead
compared to our method.
Longer history produces higher overhead.
\end{description}


\section{Related Work}\label{sec:relw}
Boothe\cite{boothe-bdb} made a C debugger with
reverse execution capability
using a step counter which counts the number of step executions
and re-execution from the beginning of debuggees.
The capability could be also implemented
with our timestamp counter and re-execution.
The difference comes from the purpose of each project.
Boothe made reverse execution version of existing debugger commands
such as ``backward step'', ``backward finish'', and so on.
Since we try to implement more abstract control of program execution
than raw debugger commands,
the counter of step execution is too expensive for our purpose.

Feldman et al.\cite{igor}, Moher\cite{provide}
and Wilson et al.\cite{demonic}
save complete memory history of process
to achieve fully random accessibility to program states.
Their systems have to deal with large ``log''.
Our system, however, saves only a pair of line number and
value of timestamp to obtain the same capability
by assuming the determinism of debuggees.

Lieberman et al.\cite{zstep,ecoop87} developed a reversible,
animated source code stepper, ZStep95.
Its modified interpreter saves the order
of evaluating S-expression of Lisp programs
to provide fully reversible execution.
ZStep95 also provides correspondence between
a S-expression and a graphical output
which is produced by the expression.
The correspondence is similar to position,
but the interpreter supports only a subset of Lisp
and works very slow.

Bertot\cite{occurrences} introduced ``occurrences'' into
the lazy $\lambda$-calculus,
which makes copies of a subtree in reduction.
For example, when an expression $e$ is applied to a lambda function
$\lambda x.x + x$, two copies of $e$ will be made and used for
both operands of $+$.
This can be regarded as creation of multiple positions
from one correspondent location at the time of execution.
Bertot achieved a breakpoint capability,
which is set at an expression and enables the program to stop
at any copy of the expression is evaluated.
In procedural languages, the identification is very easy;
use the address of the instruction as a breakpoint.
Their purpose is to unify multiple positions to the location;
our purpose is to distinguish positions from each other.

Zeller et al.\cite{ddmin,cechain} propose
``Delta Debugging'' which automatically find out data or variables
which are concerned with errors
by comparing the input data or variables
which exit normally with ones which cause errors.
It is very useful method
and similar with our method in terms of
reducing the labor of programmers by power of recent computers.
We, however, want to establish
more interactive and flexible debugging method
and our method is complementary to their one.

Ducass\'{e}\cite{coca-icse} allows the programmer
to control the execution
not by source statement orientation,
but by event orientation
such as assignments, function calls, loops, and so on.
Users write Prolog-like forms
to designate breakpoints which have complex conditions.
This mechanism is complementary to our system
and suitable for a front end of it
in order to designate appropriate positions
where we would move control point to.

Templer et al.\cite{cci,cci-light}
developed a event-based instrumentation tool, CCI,
which inserts instrumentation codes into C source codes.
The converted codes have platform independence.
The execution slowdown, however, is
2.09 times in the case of \texttt{laplace.c}
and 5.85 times in the case of \texttt{life.c}\cite{cci-light}.
In order to achieve position system,
events only about control flow should be generated.

Larus et al.\cite{eel} made EEL,
which is a library for building tools to
analyze and modify an executable program.
Using EEL, we could implement the insertion of
codes to maintain timestamp in executable code level.
The solution, however, is dependent on a specified platform,
so we chose the intermediate code level and
modified GCC.

Binder et al.\cite{prcjava}
integrated a resource management system of CPU and memory
into J-SEAL2 mobile agent system
using bytecode transformation for complete portability.
They use a counter which counts statements executed
for CPU resource management and
each thread executed updates the counter at each basic block.
They reduce the frequency of the update using control flow analysis
and the overhead of the system including other components
of agent system is $1.41$ times of slowing down in the worst case.
Hayami et al.\cite{bctrans-javacpu}
also inserted similar codes of counter update
using bytecode transformation for the same purpose.
They implemented more fine grain management
and the overhead is $1.63$ times of slowing down.
These results are better than ours,
because our implementation uses method invocation
and its overhead is serious, we think.


\section{Conclusion}\label{sec:conc}
We proposed a new idea, position,
as the base of execution control.
It introduces an absolute coordinate into program traces
and indicates a point in traces.
In order to implement it,
we introduced a counter, timestamp,
as a global variable of the debuggee,
which increases whenever the control point jumps backward.
Position is expressed
as a pair of the line number and the timestamp value.
We introduced the idea of dynamic breakpoint
as ``breakpoint at a position''
and described three applications.

We also described the implementation details of
the timestamp system for C and Java programs
using the modification to GCC and the bytecode transformation respectively.
Our GCC and bytecode transformer insert the codes of increment of timestamp
at certain locations which cause control jumps.
We measured the result of runtime overhead of C and Java implementations
and increase of size of transformed class files
and showed that they are acceptable for real use.


\section{Future Work}\label{sec:futu}
The bytecode transformer transforms all the methods in given class files.
We should make it able to do selectively.

The driver program of reverse watchpoint
for deterministic C programs is completed
but one for Java, based on JDB, is still under development.
Although JDB included in JDK-1.4.1 does not provide
the support of ``watchpoint to individual objects'',
WatchpointRequest class of JPDA in JDK-1.4.1
now have the capability of adding instance filters
by using the \texttt{addInstanceFilter} method.
JDB will have the support soon.

For non-deterministic programs,
the applications described in section \ref{subsec:pos-app}
do not work well without appropriate replay mechanisms.
If non-determinism of debuggees was based on
external environment such as input data,
we could use some tools to record and replay the environment.
For example, Xlab\cite{xlab} could be used
to record and replay X window system events.
If debuggees have internal non-determinism
such as multi-threaded programs,
tools to replay the timing of thread switchings
would be needed.
Choi et al.\cite{dejavu} implemented a modified Java VM
which can replay multi-threaded Java programs.
There is another way that the timing is saved via JPDA.


\section{Acknowledgments}
Thanks to Naoshi Higuchi for the wealth of his knowledge
about Java programming and its APIs.


\bibliography{bibliography-e}

\end{document}